\documentclass{aa}
\usepackage{graphicx}
\usepackage{txfonts}
\usepackage{color} 

\begin{document} 

\title{The solar survey at Pic du Midi: calibrated data and improved images}

\author{Laurent~Koechlin  \inst{1}, 
   Luc Dettwiller  \inst{2}, 
   Maurice Audejean \inst{3}, 
   Maël Valais  \inst{1}, 
    Arturo L\'opez Ariste  \inst{1} }
    
\titlerunning{The solar survey at Pic du Midi}
\authorrunning{Koechlin et al.}

\institute {Institut de recherches en astrophysique et planétologie,
 Université de Toulouse, CNRS, CNES, UPS, 14 avenue Edouard Belin, 31400 Toulouse, France
\and
Lycée Blaise Pascal, 36 avenue Carnot, 63037 Clermont-Ferrand cedex, France
\and
Observateurs associés, rue de la Cau, 65200 Bagnères de Bigorre, France}

  \abstract
   {At Pic du Midi observatory we carry out a solar survey with images of the photosphere, prominences and corona. This survey, named CLIMSO (
   {\it  CLIchés Multiples du SOleil}), is in the following spectral lines: Fe~XIII corona (1.075~$\mu$m), H$\alpha$  (656.3~nm) and He~I (1.083~$\mu$m) prominences, H$\alpha$ and Ca~II (393.4~nm) photosphere. All frames cover 1.3 times the diameter of the Sun with an angular resolution approaching one arc second. The frame rate is one per minute per channel (weather permitting) for the prominences and chromosphere, and one per hour for the Fe XIII corona. 
This survey started in 2007 for the disk and prominences, and in 2015 for the corona. We have almost completed one solar cycle, and hope to cover several more, keeping the same wavelengths or adding others.}
   {Make the CLIMSO images easier to use and more profitable for the scientific community.}
   {Providing ``science-ready'' data. We have improved the contrast capabilities of our coronagraphs, which now provide images of the Fe~XIII corona, in addition to the previous spectral channels. We have also implemented an autoguiding system based on a diffractive Fresnel array for precise positioning of the Sun behind the coronagraphic masks.}
   {The data (images and films) are publicly available and downloadable through virtual observatories and dedicated sites: {\it e.g.} http://climso.irap.omp.eu. For the H$\alpha$ and and  \ion {Ca} {II}  channels we calibrate the data into physical units, independent of atmospheric or instrumental conditions: we provide solar maps of spectral radiances in $\rm {W m^{-2} sr^{-1} nm^{-1} }$. The instrumental improvements and the calibration process are presented in this paper. }
{}

   \maketitle
\section {Introduction}

Pic du Midi de Bigorre in the French Pyrenees is the observatory where Bernard Lyot invented coronagraphy and obtained remarkable images in the 1930s and 1940s, see for example \cite{Lyot1930} \cite{Lyot1945} \cite{Lyot1950}. 
Since then, Pic du Midi has regularly provided high quality solar and coronagraphic data. 

In a Lyot coronagraph, the bright image of the solar photosphere is blocked by an occulting disk of same diameter in the focal plane. Furthermore, the light diffracted around the occulting disk, which would be too bright to allow observing the corona, is in turn blocked by a ``Lyot stop'' in a pupil plane. Such an optical setup associated with precise baffling allows for a high rejection factor of the solar light in the surrounding field, also called here {\it dynamic range}, reaching $10^6$.

\subsection {Scientific objectives}
 \label{sec:objectives}

The survey presented here aims to help the scientific community for solar studies by providing a large set of solar images and films with dense temporal sampling: 1 minute, and long span: several solar cycles, at stable observation conditions (field, wavelengths and bandpass, arc-second angular resolution, high contrast). 

\begin{figure*}
 \centering
\includegraphics[width=.489\linewidth] {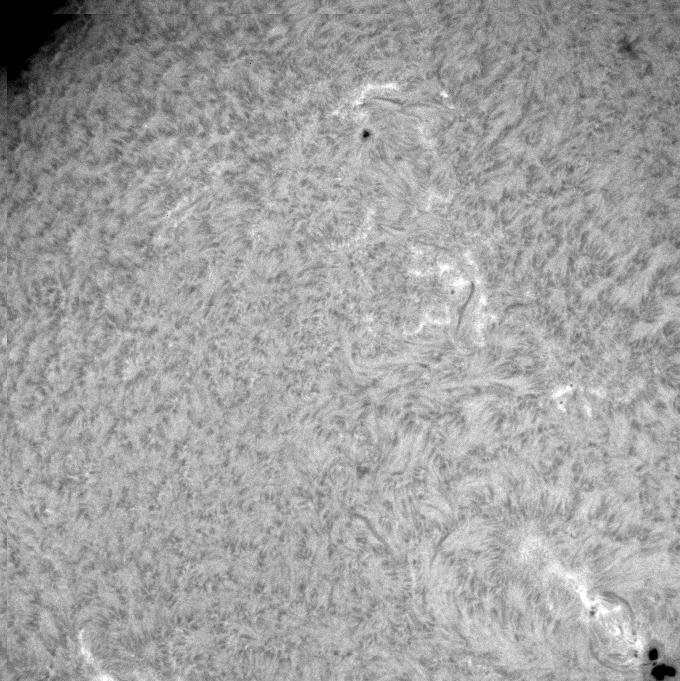}
\includegraphics[width=.489\linewidth] {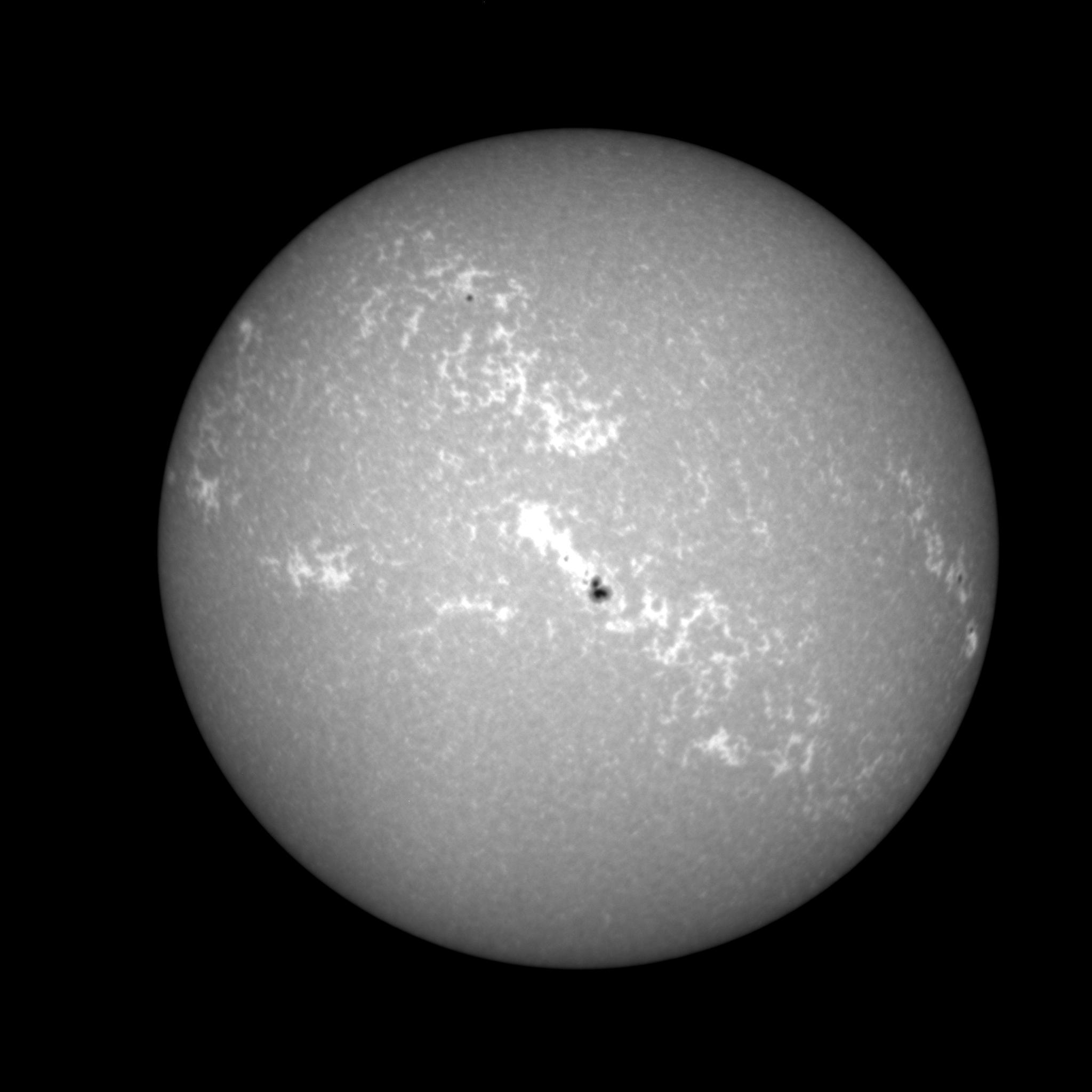} 
\caption{ {\bf Left:} Solar chromosphere in H$\alpha$ at 656~nm. 
This cropped frame contains the upper left quarter of the solar disk: 
approximately 810x810~arc-seconds, 
compare with the simultaneous image in Ca~II on the right. The original image covers the full disk in 2k*2k pixels. Images are taken every minute, from one hour after sunrise to one hour before sunset, weather allowing.
{\bf Right:} Solar chromosphere in the Ca~II line at 393.4~nm. These two images were taken with $l_1$ and $l_2$ at Pic du Midi, 2015-03-27.}
\label{fig:disk_Halpha+CaII}
\end{figure*}

\begin{figure*}
 \centering
\includegraphics[width=.486\linewidth] {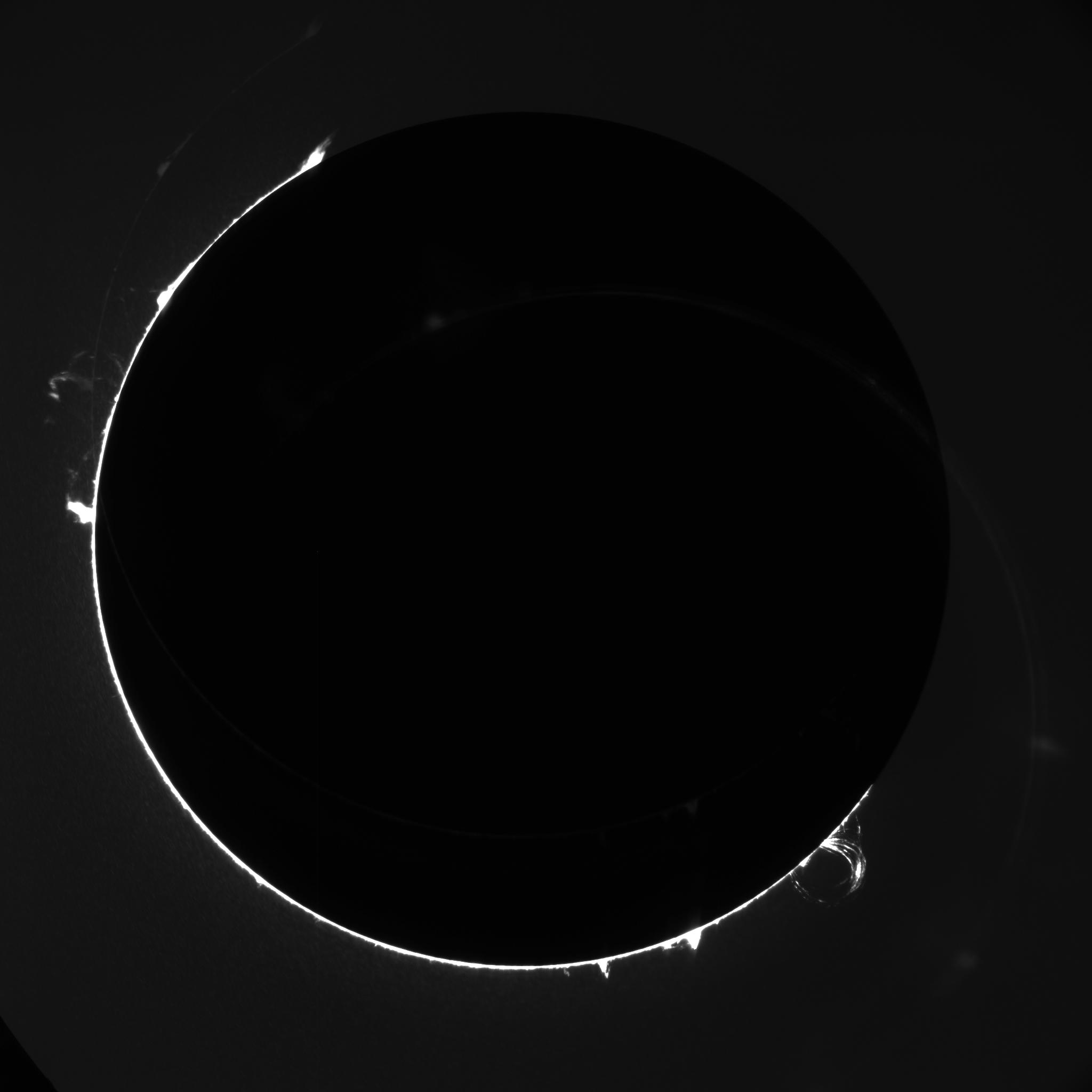} 
\includegraphics[width=.486\linewidth] {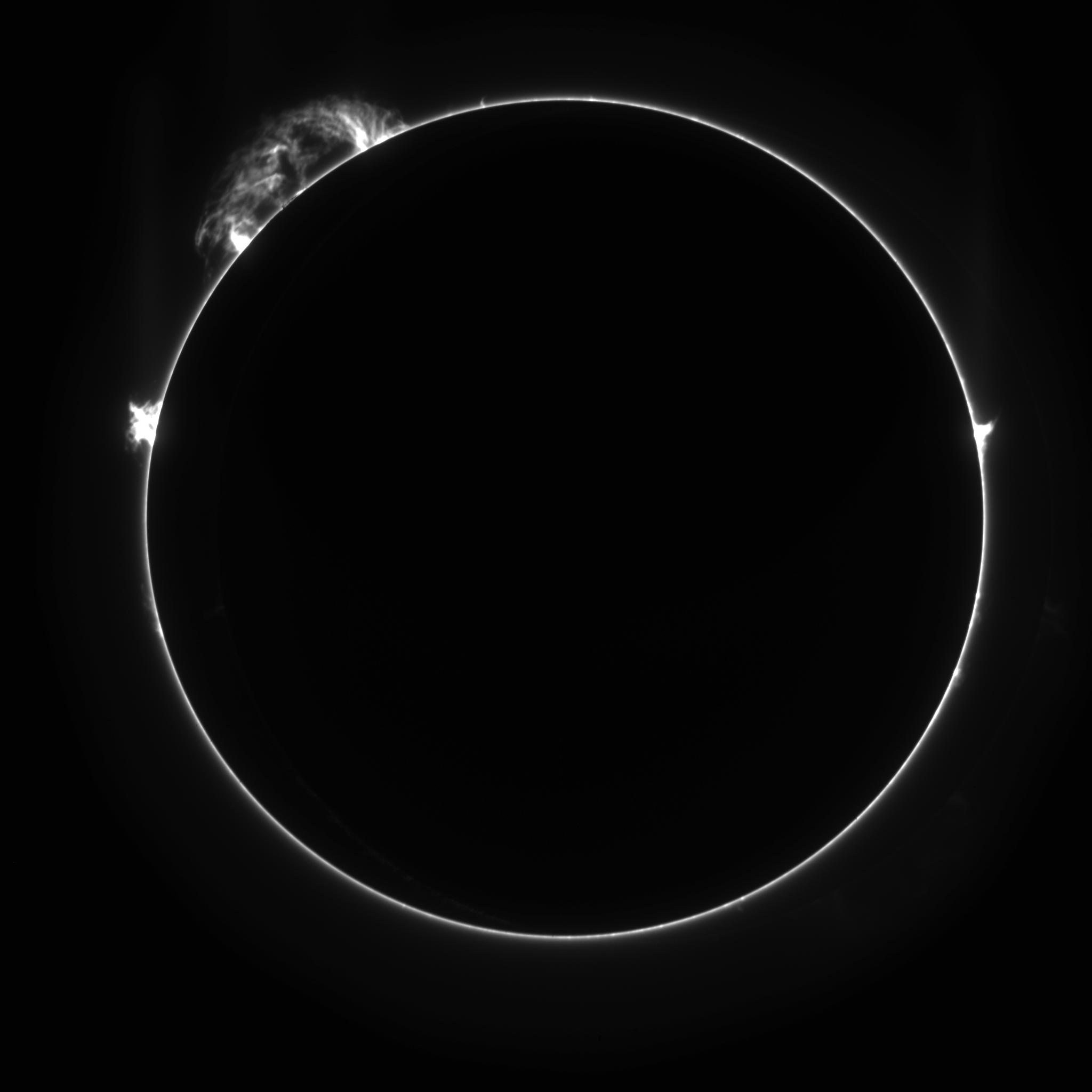} 
\caption{ {\bf Left:} Solar prominences in H$\alpha$ at 656.28~nm, taken with $c_1$ during the solar eclipse of 2015-03-20. The dark lunar disk blocks the light in the upper right part of the field.
{\bf Right:} Solar prominences in the \ion {He} {I} line at 1.0830 $\mu$m, taken with $c_2$, 2015-03-27. }
\label{fig:Halpha+HeI}
\end{figure*}

We keep the same observation conditions all through the survey, except for improvements in image quality, such as in figs.\ref{fig:disk_Halpha+CaII} and \ref{fig:Halpha+HeI}.

At present, five instruments contribute to the CLIMSO solar survey: two coronagraphs, two solar telescopes: (\cite{Dettwiller2008}), and an autoguider. The autoguider uses a diffractive Fresnel array as a focuser and provides images of the Sun, in order to center the other instruments. 
CLIMSO is an acronym for either {\it CLIchés Multiples du SOleil}  or  {\it Christian Latouche IMageur SOlaire}.

We have been carrying out this survey since 2007, thanks to the important contribution of the "Observateurs associés" volunteer astronomers (www.climso.fr) who contribute in many aspects: instrumental and software development, financial support and, last but not least, year-long handling of image acquisition.

\subsection {The instruments set}
 \label{sec:instrum}

Five instruments are placed on a single equatorial mount:
\begin {itemize}
\item $c_1$:  H$\alpha$ coronagraph, $\oslash \, 20 \, {\rm cm}, \, \lambda \, 656.3\, {\rm nm}, \,  \Delta\lambda \, 250\, {\rm pm}$;
\item $c_2$:  He~I coronagraph,  $\oslash \, 20 \, {\rm cm}, \, \lambda \, 1.083\, {\rm \mu m}, \,  \Delta\lambda \, 250\, {\rm pm}$;  
\item on the same  $c_2$: Fe~XIII channel $  \lambda \, 1.0747  {\rm \mu m},  \,  \Delta\lambda \, 250\, {\rm pm}$; 
\item $l_1$: H$\alpha$ telescope,   $\oslash \, 15 \, {\rm cm}, \, \lambda \, 656.3\, {\rm nm}, \,  \Delta\lambda \, 50\, {\rm pm}$;
\item $l_2$: Ca~II telescope,  $\oslash \, 9 \, {\rm cm}, \, \lambda \, 393.4\, {\rm nm}, \,  \Delta\lambda \, 250\, {\rm pm}$;
\item and a Fresnel diffractive telescope that feeds the autoguider: $\oslash \, 6.2 \, {\rm cm}, \, \lambda \, 632.8\, {\rm nm}, \,  \Delta\lambda \, 1\, {\rm nm}$.
\end{itemize}
The four solar telescopes with their acquisition software were conceived and built in 2006 by J.-C.~Noëns, L.~Dettwiller, D.~Romeuf and subcontractors: (\cite{Dettwiller2008}).
Since then, they have been providing images and films on a regular basis to the CLIMSO data base http://climso.irap.omp.eu.

Among several upgrades, solar telescope $l_1$ now has high quality optics and a new H$\alpha$ filter. We have improved the baffling and especially the correction of filters leakage in coronagraph $c_2$.This led in 2015 to the $10^6$ dynamic range allowing for images of the solar corona.

Here we express here the ``dynamic range'' of a coronagraph as the ratio between the brightness in the image of the photosphere and the residual light in the surrounding field, at the final focal plane. This could also be called "rejection factor".

In good weather conditions, a set of four images is taken per minute, and one of the solar corona in the Fe~XIII line at 1074.7~nm per hour. All images are uploaded to the database each night, along with several films made from them. During daytime, a few images are selected for real-time uploading to the data base.


\section {Guiding system}
 \label{sec:iguidage}

The fifth telescope is used for guiding. Its main optics consist of a Fresnel diffractive array, 2.5~m in focal length at $\lambda = 633$~nm, placed close to the $c_1$ coronagraph. The coronagraph and diffractive optics for the guider are fastened to each other, thus limiting differential flexion.

\begin{figure*}
\centering
\includegraphics[width=.24\linewidth] {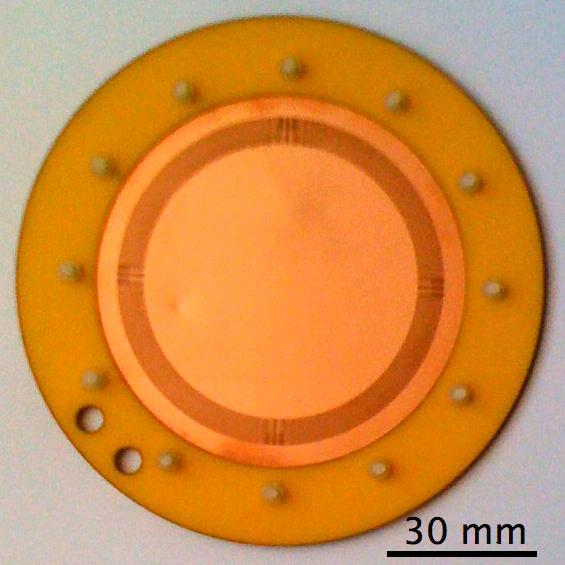} 
\includegraphics[width=.24\linewidth] {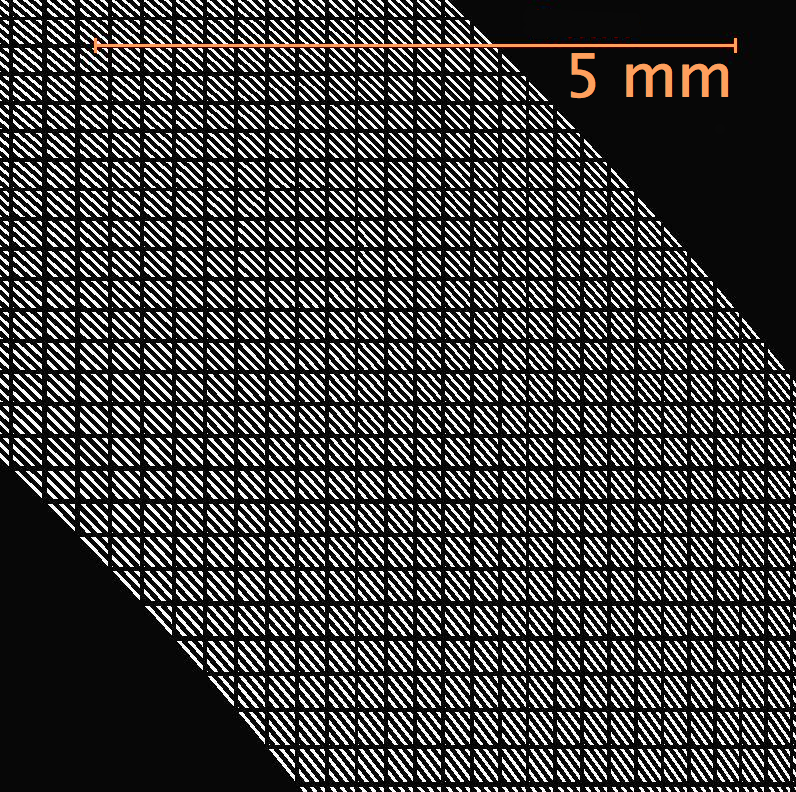} 
\includegraphics[width=.24\linewidth] {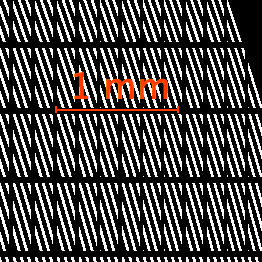}
\includegraphics[width=.24\linewidth] {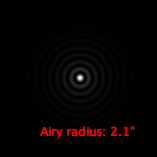}
\includegraphics[width=.485\linewidth] {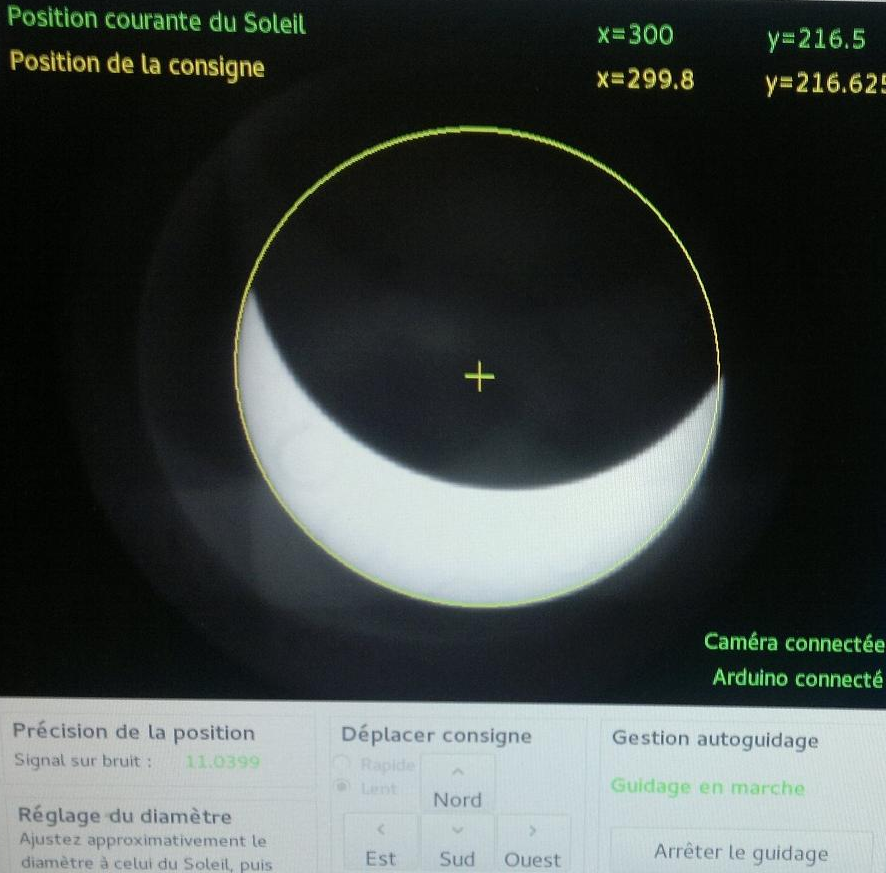} 
\caption{ The autoguider. Its objective lens is a ring-shaped Fresnel array (top left): a thin copper plate with several thousand subapertures. An image of the Sun is formed by diffraction and interference through the holes. Top center: two enlarged views of the Fresnel array holes. Top right: image of a monochromatic point source (PSF) given by the Fresnel array. Bottom: control screen {\it (photo by Géraldine Pedezert)} of the autoguider during the solar eclipse, 2015-03-20. The solar disk is correctly located, even though only a crescent remains visible. }
\label{fig:grilleFresnel}
\end{figure*}

The Fresnel diffractive array shown in Fig.\ref{fig:grilleFresnel} is a copper foil 80~$\mu$m thick and 65~mm in diameter with several thousand holes, based on the diffraction principle: (\cite{Fresnel1818}, \cite{Soret1875}). It acts as a Fresnel zone plate and forms a sharp image of the Sun on a camera, which sends 1 frame/s to an image processor. The diffractive Fresnel array has a binary transmission (\cite{Koechlin2005}): there is no optical material in this objective `lens', only opaque material (here we use a thin copper plate) with specially shaped holes that focus light by constructive interference. This setup is lightweight, steady, and affordable. 

This optical concept provides diffraction-limited images. It works with broadband (non monochromatic) light as it can be corrected from chromatic aberration: Fresnel arrays have given high contrast images of sky objects, see \cite{Koechlin2012}, \cite{Koechlin2014}. Although they refer to tests with $20\times20\, {\rm cm}$ square Fresnel arrays, these two publications provide references of studies for larger diffractive arrays.

Diffraction focusing causes chromatism in the resulting images at prime focus, but here we do not need to treat it: we just put a narrowband filter ($\lambda = 633$~nm, $\Delta \lambda= 1$~nm) near focus. This wavelength of He-Ne lasers does not correspond to a feature in the solar spectrum, so the images of the solar disk in the autoguider are prominence-free. 
This He-Ne filter is affordable; its spectral finesse (1~nm) is slightly too low for complete chromatism cancelation in this optical configuration, but the images are almost diffraction-limited.

To achieve a good contrast we have to block the light from unwanted diffraction orders—mainly order zero passing as a plane wave, unfocused through the Fresnel array. To block it we use a central obscuration, casting its shadow on the focal plane. By geometric design, this also removes all positive and negative diffraction orders except the desired one. As the Sun is not a point source, the central obscuration (here 52 mm) needs to be larger than the required field for guiding. On the other hand, the outer diameter of the objective is limited by the size of the narrowest Fresnel rings in the diffractive array that can be grooved, itself limited by the waist of the UV laser beam used to manufacture it. The local machine tools that were used have a laser beam width of 20 $\mu$m, thus limiting the Fresnel array diameter to 62 mm for 2.5~m focal length at 633 nm. Recent machine tools with thinner laser beams can manufacture much wider Fresnel arrays with equal focal length.  

\begin{figure*}
\centering
\includegraphics[width=.486\linewidth] {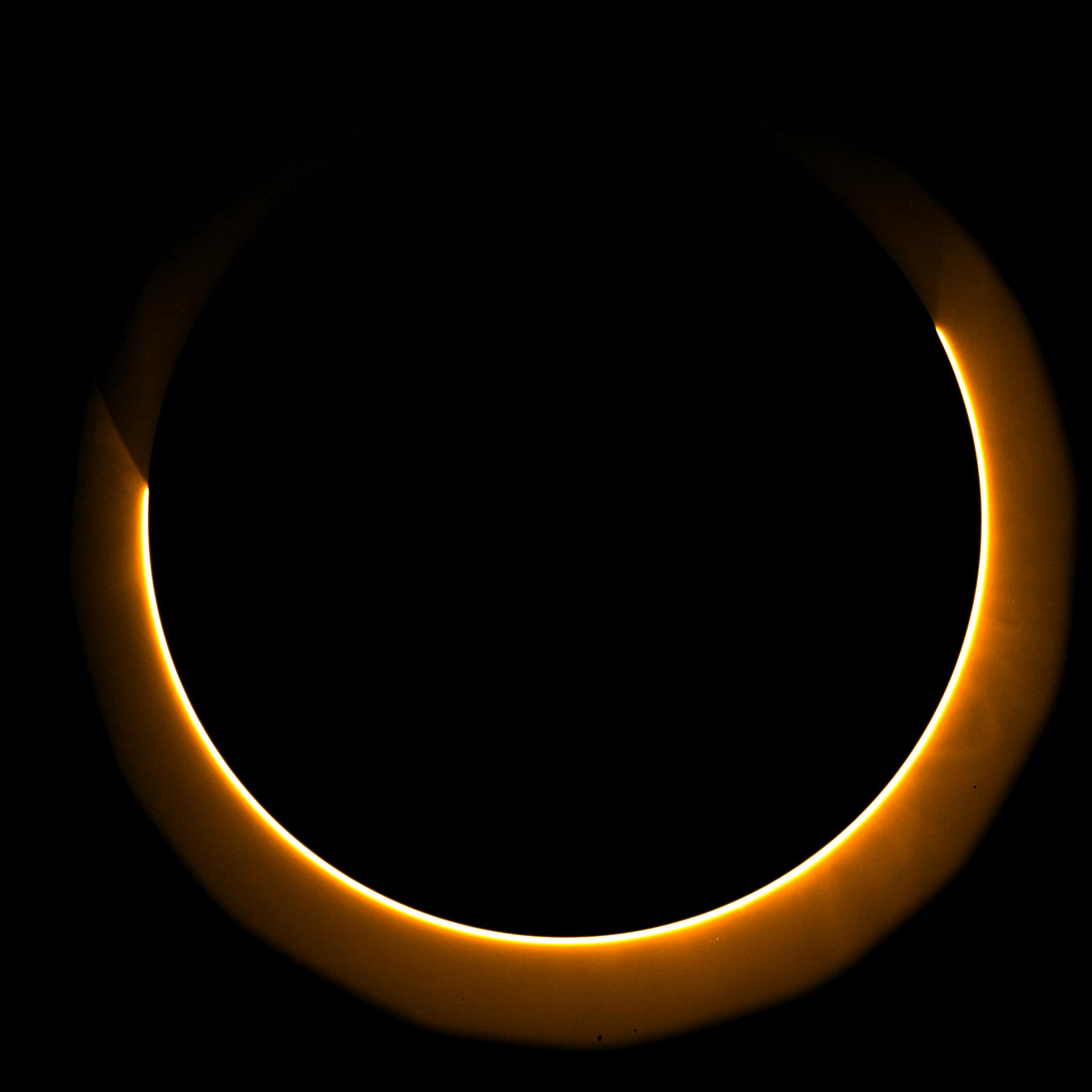} 
\includegraphics[width=.487\linewidth] {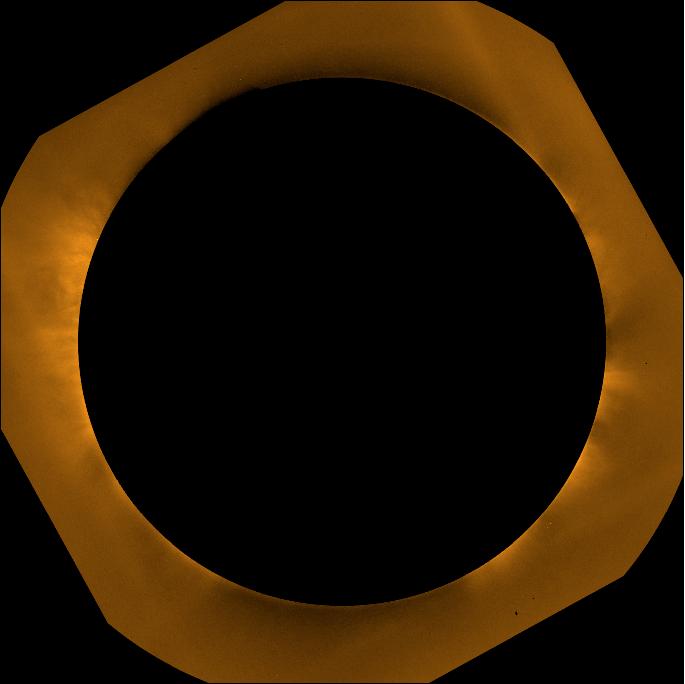} 
\caption{ {\bf Left:} one of our first images of the Fe XIII corona at 1075 nm, taken during a partial solar eclipse, 2015-03-20. One can see a coronal loop on the right, coronal jets, and the dark lunar disk blocking the upper part of the field.
{\bf Right:} a more recent image of the solar corona at 1075 nm, a composite of 5 images taken between 07:39 and 07:57 on 2015-04-12, 6s exposure time each. Images of the Fe XIII corona are now made on a regular basis at intervals of 30 minutes every day, weather allowing.}
\label{fig:couronne_FeXIII}
\end{figure*}

An image of the Sun is taken every second by a SBIG-8300F camera at focus (behind a small lens adapting the field scale to the camera sensor), then sent to the autoguider software which keeps the five instruments precisely pointed, even when the Sun disk is partly blocked by clouds or by the moon, see Fig.~\ref{fig:grilleFresnel}, bottom.

The solar disk is processed to extract its edge (ring or arc), which in turn is correlated with a reference circle. The guiding software adapts the diameter of that circle to the apparent diameter of the Sun. The position of the reference circle can be either manually or automatically adjusted to any desired position. To reduce the processing time we do not use FFTs but sparse matrix algorithms. 
The computed distance between the center of the reference circle and the center of the solar disk is used to determine a command sent to the equatorial mount.
The source code in C++ and a detailed documentation of the autoguider are available on github : https://github.com/maelvalais/climso-auto.

The guiding precision is better than 1 arc-second in good seeing conditions. Of course, due to the slow sampling rate of 1~s and the inertia of the equatorial mount, most of the tip-tilt agitation due to atmospheric seeing is not corrected. This limitation in performances is significant in bad seeing or high wind conditions. We plan to improve the algorithm that computes the commands to the equatorial mount based on the history of solar positions found; at present it is just a proportional response.

\section {Photometric image calibration}
 \label{sec:calib}
\subsection {Principle}
 \label{ssec:Principle}

Our photometric calibration is based on the fact that in an image of the Sun made from the ground, the sum of all pixels in the solar disk represents the total light received from the Sun in the instrument's spectral bandpass. If we have an independent measure of the corresponding solar power above the atmosphere, such as a calibrated solar spectrum obtained by a space-borne system, and if we make the approximation that the solar absolute magnitude can be considered constant, we can deduce a calibration factor for any image made under the atmosphere. This calibration globally takes into account the atmospheric and instrumental transmission for that particular image.

\subsection {Preprocessing}
 \label{ssec:Preprocessing}
In the first steps we correct the images from dark and flat-field effects induced by the cameras: CCD on solar telescopes $l1$ and $l2$, and CMOS on coronagraphs $c1$ and $c2$. As the exposure times are all very short, we consider that the bias and dark are equivalent. The pixel scale is calibrated from the angular diameter of the Sun each time a modification is made to the optics. The rotation angle of the cameras and its variation during the day is also calibrated regularly, and the images are de-rotated so that the North solar pole is up. 

The geometrical distortion is not corrected but very low: the optics in the coronagraphs and solar telescopes all have a long f-ratio and they work at small angles.

For $l1$ and $l2$ we perform a photometric calibration explained in the following. The result is a spectral radiance map of the Sun. 

At this point the cameras provide raw images of the solar disk that represent its local radiance, but they are affected by: 
\begin {itemize}
\item the Sun--Earth distance;
\item the absorption due to airmass and atmospheric conditions;
\item the instrument: optics transmission and camera response.
\end{itemize}
We calibrate those three globally, to obtain calibrated images that represent physical conditions at the surface of the Sun or its vicinity, so the pixel ``brightness'' values are in physical units. 
The choice of a relevant unit is not simple since the H$\alpha$, \ion{He}{I}, \ion{Ca}{II}~K and \ion{Fe}{XIII} light forming the images is emitted from volumes of solar plasma (chromosphere, prominences, corona) with various optical thickness and altitude. 
\subsection {Photometric units}
 \label{ssec:notions}
To make the following description of our calibration procedure as clear as possible, let us recall a few photometric definitions that will be used.

\noindent  {\it In emission}, we use the {\bf radiance} $L(x, y, \alpha, \delta)$, expressed in ${\rm W m^{-2} sr^{-1}}$, which is the flux emitted:\\
 -- in a given direction $(\alpha, \delta)$ {\it per unit of solid angle} (sr) -- here it corresponds to the line of sight; \\
 -- from a point $(x,y)$ of a surface, {\it per unit area} ($\rm m^2$) projected on a plane (here the sky plane) normal to the considered direction (the line of sight).\\

We also use the spectral radiance $L(x, y, \alpha, \delta, \lambda)$ in  $\rm  W m^{-2} sr^{-1} nm^{-1}$. 

We consider the emission towards the Earth, i.e. the emission quasi perpendicular to an emissive surface {\it in the plane of the sky}, but not necessarily parallel to the local solar surface. Thus we can express the spectral radiance in our direction $(\alpha, \delta)$ as 
$L(x, y, \lambda)$.\\
  
\noindent  {\it In reception},
{\bf the illuminance} $I(x,y) $ is the flux received:\\
-- from all directions; \\ 
-- at a given point $(x,y)$ of a collecting surface, per unit area.

\noindent {\bf The irradiance} corresponds to the illuminance when the collecting surface is perpendicular to mean propagation, which is our case. 
The irradiance $I(x,y) $ is expressed in $\rm W m^{-2}$, and the spectral irradiance $I(x,y, \lambda) $ in $\rm  W m^{-2} nm^{-1} $. \\

In order to link the radiance $L$ emitted from a point $S$ above the Sun to the radiance $L'$ received on Earth at $P$, we use Clausius' theorem \cite{Born1980},  \cite{Klein1970}, stating: 
\begin{equation}
L'= TL
\label{equ:Clausius1}
\end{equation}
where $T$ is the transmission factor associated to the beam, when we neglect the effect of diffusion on the optical path, and assume the refraction index $n\simeq1$ at $S$ and $P$. For simplicity, when $P$ is the image of $S$ by an instrument, we use the same notation $(x,y)$ for coordinates of $S$ and $P$.
As briefly explained in the first paragraph of this section \ref{sec:calib}, to obtain a radiance map of the Sun through the transmission of a telescope, our basic idea is double: first, the value of a pixel in a Lambertian situation is proportional to the radiance of the corresponding element of the Sun; second, the proportionality factor is uniform over the whole solar disk. Hence, the relative value of a pixel compared to the average all over the Sun is also the relative radiance, compared to the mean radiance of the whole Sun which is known from airborne measurements.

\subsection {Photometric calibration}
 \label{ssec:calib_1AU}

\begin{figure}
 \centering
\includegraphics[width=\linewidth] {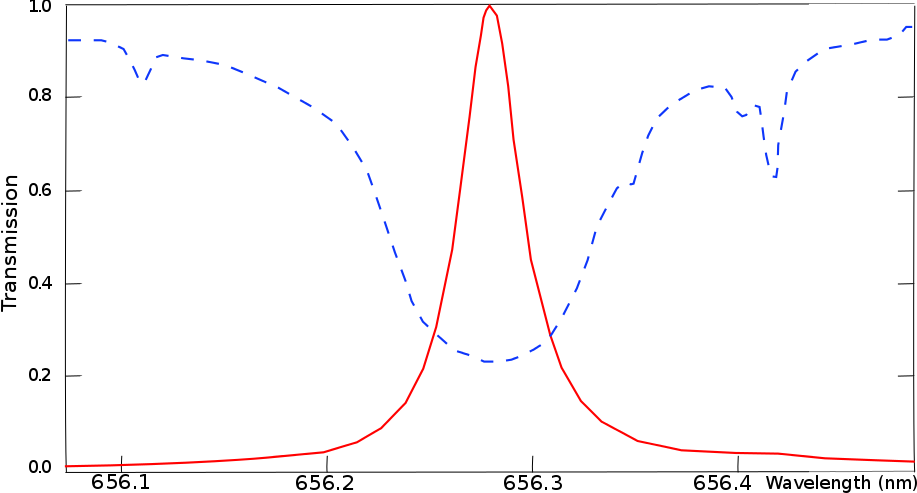} 
\caption{
{
Measurements by J.-M. Malherbe.
{\bf Red curve:} transmission profile of his $\Delta \lambda$= 40 pm bandwidth filter.
For our calibrations we have enlarged this profile to match the $\Delta \lambda$= 50 pm bandwidth profile of our filter from the same manufacturer: Daystar. We convolve the solar spectrum with this profile. The dotted line represents the central part of the H$\alpha$ line in the solar spectrum, not normalized in this figure.
}
}
\label{fig:raieHalpha}
\end{figure}

{In order to have a calibrated image of the solar disk, the physical data we want for a pixel is  the radiance towards the Earth: $L(x,y)$ of the corresponding point on the solar disk, through the normalized transmission function of our filter $T(\lambda)$ (whose maximum is $T^{max}$), plotted in Fig.\ref{fig:raieHalpha}: 
\begin{equation}
L(x,y) := \int _0^ \infty  { {T (\lambda) \over T^{max} } L(x, y, \lambda) } d \lambda
\end{equation}
$L(x, y, \lambda)$ is the local spectral radiance on the Sun (the plane in which this is defined will be discussed later).
}

{
We make the following approximations:
\begin{itemize}
\item  On the image plane we are in a Lambertian situation, close to the optical axis;
\item  $T (\lambda)$ does not depend on $(x,y)$, {\it i.e.} the filter is uniform in the field;
\item  the instrument and camera defects are flat-fielded out, but in addition the atmospheric transmission is uniform over the image field. That is valid in clear sky conditions, and can be an approximation for a uniform haze or cirrus;
\item  our cameras have a linear and uniform response (once corrected from dark, and flat-field); we assume that a raw pixel value $v(x,y)$, measured in arbitrary units `{\it adu}' for a point $P(x,y)$ in the image, is proportional to the local irradiance $L'(x, y)$ in that image.
\end{itemize}
}

{
The raw pixel value  $v(x,y) $ is given by:
\begin{equation}
v(x,y) \cong C\int _0^ \infty {T (\lambda) L(x, y, \lambda) } d \lambda
\end{equation}
\begin{itemize}
\item $C$ is a transmission factor (constant over the image field, but variable in time) driven by the instrument, the Sun-Earth geometry and the atmosphere, which will be canceled out during the calibration process.
\item The spectral transmissions of the atmosphere and optics, and the camera response, contribute to $C$ but they are slowly wavelength-dependent compared to the narrow filter bandpass $T (\lambda)$, so $C$ can be considered wavelength independent in the relevant spectral band.
\end{itemize}
}
The variations of $C$ with time, caused by changes of air mass or haziness, are taken into account by the calibration process, as shown below. Changes in the instrument, such as aging of filters or cameras, which could affect the global response, are also calibrated out by the same process. Furthermore, it can also be applied retrospectively to data that had not been calibrated when it was acquired. Care should be taken, though, that filter wavelengths are controlled and remain correct.

\begin{figure}
 \centering
\includegraphics[width=\linewidth] {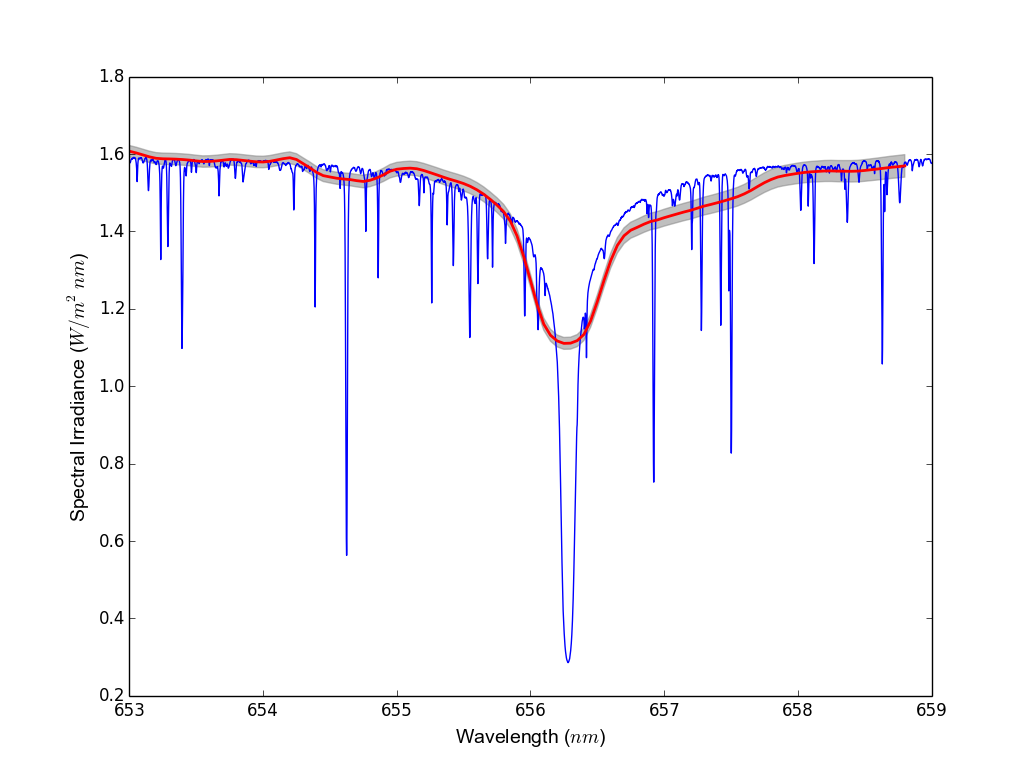} 
\caption {
{
Solar irradiance above the atmosphere in  $\rm  W m^{-2} nm^{-1}$ as a function of wavelength in nm. The {\bf blue curve} is the FTS high resolution solar spectrum near H$\alpha$,  \cite{Kurucz1984} that we use to calibrate the images taken with $l_1$ ($\lambda = 656.28$~nm; $\Delta \lambda = 50$~pm). 
The {\bf red curve} is here for comparison. It is a calibrated solar spectrum taken at the ISS by \cite{Meftah2017}. The grey zone around the red curve represents the area covered by its error bars. The wavelength axes of these two curves have been adjusted to each other. The superposition of the red and blue curves indicates a good agreement between the ISS data and our calibration of the high resolution spectrum.
}
}
\label{fig:superp_spectres}
\end{figure}

We know the equivalent bandpass $\Delta \lambda$ from the transmission function $T$, 
{
here for $l_1$: }
\begin{equation}
\Delta \lambda := \int_0^ \infty {T(\lambda) \over T^{max} }d\lambda \cong (0.05\pm  {0.005}  ) \rm \,nm.
\end{equation}
%
{
The  filter bandwidth and its precision are given by the manufacturer (Daystar).
}

We take, from the literature and documentation, the following data:
\begin{itemize}
\item standard solar spectrum $I_{\lambda,m}$ (averaged over a solar cycle), itself implicitly based on the solar `constant', {\it i.e.} the standard solar irradiance at 1 AU;
\item spectral transmission of our filter: $T(\lambda)$.
\end{itemize}
As we use narrowband filters, we need a calibrated standard solar spectrum at high spectral resolution. We start from an uncalibrated standard spectrum at high resolution: the Meudon Solar Atlas, Delbouille et al. (http://bass2000.obspm.fr/solar-spect.php), then calibrate it with a calibrated standard solar spectrum at lower resolution: rredc.nrel.gov/solar/spectra/am1.5/ASTMG173/ASTMG173.html
From that we compute $I_m $ the standard solar irradiance above the Earth atmosphere integrated through the spectral profile of our filter, 
{ 
and we find its numerical value for $l_1$ (also in table \ref{tab:numval}):
}
\begin{equation}
I_m :=    \int _0^ \infty {  {T (\lambda) \over  T^{max} } I_{\lambda,m}(\lambda)  d\lambda \cong (0.018\pm0.001)  \rm W m^{-2} }  .
\label{eq:I_m}
\end{equation}
{ 
The $\pm0.001$ error bar comes from the solar spectrum calibrations and the filter spectral curve: we have numerically integrated equation (\ref{eq:I_m}) to obtain the numerical value in $ \rm W m^{-2}$. 
}

From this external data, and neglecting the cosine in the relation between the radiance and the elementary flux, we deduce $L_m$, the standard radiance averaged over the solar disk $(D_m)$, and integrated through the transmission of the instrument:
\begin{equation}
L_m \cong I_m \; / \; \Omega_m \cong \int _{(D_m)}L \; d^2\Omega  \; / \;  \Omega_m
\label{eq:irradiance}
\end{equation}
where $\Omega_m\cong 0.680\,\times10^{-6}\rm \,sr$ is the solid angle of the solar disk at 1~AU.

The $\pm 5$\% uncertainty resulting from our estimation of $I_m$ in equation (\ref{eq:I_m}) is much greater than the errors due to the effects of the cosine factor neglected in equation (\ref{eq:irradiance}) and of the distorsion. 

{ 
Now we compare $L_m$ with the average of the raw pixel values $v(x, y)$, over the $N$ pixels covered by the solar disk image. 
\begin{equation}
\sum_{j=1}^{N}{  {v}(x,y) \over  N } := {\bar v}   
\end{equation}
This average solar pixel value $ {\bar v}$ corresponds to the apparent brightness of the solar disk through all the instrumental chain. From an independent source we have its calibrated brightness: $L_m $, calculated with equation (\ref{eq:irradiance}), and we can deduce the calibration factor $C$ with:
\begin{equation}
{\bar v}  \cong C L_m 
\end{equation}
}
It allows to eliminate $ C$, and we finally find:
\begin{equation}
  L (x,y)  \cong   L_m \; v_r   (x,y) 
\end{equation}
where $v_r(x,y) := v(x,y)/ {\bar v}$ is the relative value of a pixel compared to the average, and
\begin{equation}
L    (x,y) \cong  
K\;  v_r(x,y) 
 \end{equation}
 with 
 $  K :=   L_m / \Delta \lambda .$
 
Similarly, for our other solar telescope $l2$ in the  \ion {Ca} {II}  line, we calibrate the images. We have a relatively inaccurate knowledge of the spectral transmission for the 10-year-old filter in telescope $l2$, in addition to the variability of the line, so we provide ``calibrated'' images with low precision. We plan to buy better filters in a few months. 

The numerical values are listed in Table \ref{tab:numval}. 
In the third and the fourth line of this table we have, corresponding to each filter, the mean spectral irradiance from the Sun above the Earth atmosphere, then this spectral irradiance integrated through the spectral curve specific to the filter.
The calibration in Ca~II is less precise because of the solar variability in that line, in addition to the aging of our Ca~II filter.
These are used to generate the calibrated images provided by the public data base: http://climso.irap.omp.eu/data. To keep a sufficient dynamic range in the 16-bit integer data, the pixel values in the data base may be adjusted by a factor of 10. This will be documented in the FITS header.
At present only the recent images have a calibrated counterpart, but all images since 2007 are going to be reprocessed and calibrated.

\begin{table}
{
\begin{center}
\caption{
Calibration factors. 
{\bf First pair of lines}: specifications of our filters. 
{\bf Middle}:  irradiance from the calibrated solar spectrum in blue Fig.\ref{fig:superp_spectres}, integrated  over the transmission curve. 
{\bf Bottom}: calibration factors.
\label{tab:numval}  }
\begin{tabular} {ccc}
\hline 
\hline Channel & H$\alpha$  ( $l_1$) &  \ion {Ca} {II}  ( $l_2$) \\
\hline central filter wavelength  $\lambda$  ($nm$)  & 656.28 &  393.37 \\ 
 equiv. filter bandwidth  $\Delta \lambda$  ($nm$)  & 0.050 & 0.25 \\
$ I_m  (\lambda)$ above atm.  /  W m$^{-2}$nm$^{-1}$ &  $0.36 \pm 0.02$  &  $0.33 \pm 0.07$ \\
$ I_m $ integ. through filter / W m$^{-2}$  & $0.018$ & 0.083 \\
 $K$  solar surf.  /  kW m$^{-2}$nm$^{-1}{\rm sr^{-1}}$ & $529 \pm30$ & $485 \pm90$ \\
$ L_m $  solar surf. / kW m$^{-2} {\rm sr^{-1}}$  & $ 26.5 \pm 1.5$ & $122 \pm 26$\\
\hline
\hline
\end{tabular}
\end{center}
} 
\end{table}

\subsection {Future calibration of the H$\alpha$ prominences}

The spectral bandpass of the H$\alpha$ filter in coronagraph $c_1$ is $ \Delta \lambda_{c1} = 0.25 \, {\rm nm}$. It can be considered wider than the emission lines of prominences, but narrow enough so that no significant emission from another phenomenon contributes to the pixel value
$c_1(x,y)$ and

\begin{equation}
c_1    (x,y) \cong C_{c1} \; T_{c1}^{max}  L_{\rm H \alpha} (x,y) 
\end{equation}

\noindent where $L_{\rm H \alpha} (x,y)$ is the radiance at $(x,y)$ over the H$\alpha$ line of the prominences.

The calibration of  images from  coronagraph $c_1$ in Fig.\ref{fig:Halpha+HeI} left, cannot be done similarly to $l_2$ and $l_1$ because it would require sampling the solar disk, which is too bright for $c_1$ and not adapted to its ring-shaped field. 

However, as $c_1$ and $l_1$ are centered on the same wavelength and both see the prominences (although $c_1$ is much more sensitive), a cross-calibration is possible. It is not simple to implement, but we propose the following approximations:
\begin {itemize}
\item { the images on $c_1$ and $l_1$ can be considered simultaneous (they are both taken within a few seconds interval);}
\item { the instrument characteristics and camera sensitivities evolve slowly and the average ratio between the responses of $c_1$ and $l_1$ can be considered constant for a few hours.}
\end {itemize}

This image calibration for $c_1$ is not done yet, but planned for the near future:
we will use sets of prominences having brightnesses within the linear response of both instruments $l_1$ and $c_1$: we find by least square fit the ratio ${ Pro (l_1)  / Pro (c_1)}$ between images made by $ l_1$ and images by $c_1$ of the same prominences. 

As the radial velocities may drive some of the emission beyond the narrow bandpass of $l_1$ while still within the wider bandpass of $c_1$, special care will be taken in the algorithm to extract a precise brightness ratio from the histogram of ratios found.

\subsection {No calibration yet for coronagraph $c_2$, but...}
 $C_2$ looks into the He~I line at 1.0830~$\mu$m: Fig.\ref {fig:Halpha+HeI} right, and Fe~XIII at 1.0747~$\mu$m: Fig.\ref {fig:couronne_FeXIII}. It cannot be calibrated yet as we have no solar photometer at that wavelength. We could compare the optics transmission and camera response at different wavelengths, then estimate a ratio for the air-mass effect, but this would be prone to many cumulative errors. 
 For the moment and pending the implementation of a photometer in the corresponding waveband, for $c_1$ and $c_2$ we upload merely normalized images of prominences and corona.

\section {Conclusion}
We hope that our calibrated images contribute to the development of solar physics. We do not claim a high photometric precision immediately, but our calibrated images should be useful. 

This calibration uses the solar spectrum, itself based on the solar constant, hence it can be done for any image, even if obtained long ago, as long as it contains the solar disk (or a known fraction of it) with known wavelength and bandpass. To improve precision, we plan to reject spots, filaments and eruptive regions when computing the solar disk average of each image.

This CLIMSO data base was started in 2007 and we hope to carry on for several solar cycles, and to complement other ground-based or space surveys. You are welcome to use it! 

\begin{acknowledgements}
This set of instruments is operational all year long thanks to the 90 volunteer astronomers: the ``Observateurs associés", who share the tasks of image acquisition, instrument development and software, with the staff at Observatoire Midi-Pyrénées (OMP) and Institut de recherche en astrophysique et planétologie (IRAP), in scientific collaboration with the ``Programme national Soleil-Terre'' (PNST).
The Observateurs associés are funded by Fiducial (GEO Christian Latouche); 
IRAP and OMP are funded jointly by CNRS and Université de Toulouse.

Special thanks to Raphaël Jimenez, Martine Lafon, Philippe Saby, Franck Vaissière (president of the Observateurs associés),
and the OMP staff, whose work has been crucial to the good operation of these instruments, computers and networks.
We thank the anonymous referee for his/her helpful comments.
\end{acknowledgements}

\bibliographystyle{aa} 
\bibliography{calicoro.bib} 

\end{document}